\documentclass[aps, %
twocolumn, %
preprintnumbers, %
nofootinbib, %
floatfix, %
superscriptaddress, %
]{revtex4} %

\IfFileExists{srcltx.sty}{\usepackage[active]{srcltx}}

\usepackage{amssymb}%
\usepackage{amsmath}%
\usepackage[dvips]{graphicx}%
\usepackage{bm}

\usepackage{paralist} %

\let\jnlstyle=\rm \def\jref#1{{\jnlstyle#1}} 
 \def\apj{\jref{ApJ}} 
\def\apjs{\jref{ApJS}}   
\def\aap{\jref{A\&A}}   
  
 \def\mnras{\jref{MNRAS}}

\newcommand{\fov}{{\mathrm{fov}}} 
\newcommand{\xrb}{{\textsc{xrb}}} 
\newcommand{\cl}{{\mathrm{cluster}}} 
\newcommand{\coma}{\mathrm{Coma}} 
\newcommand{\virgo}{\mathrm{Virgo}} 
\newcommand{\dm}{{\textsc{dm}}} 
\newcommand{\kev}{\:\mathrm{keV}} 
\newcommand{\kpc}{\:\mathrm{kpc}} 
\newcommand{\mpc}{\:\mathrm{Mpc}} 
\newcommand{\gal}{\mathrm{MW}} 

\begin{document}
\preprint{CERN-PH-TH/2006-005}
\title{Where to find a dark matter sterile neutrino?}%
\author{A.~Boyarsky}%
\affiliation{CERN, Theory department, CH-1211 Geneve 23, Switzerland} %
\affiliation{\'Ecole Polytechnique F\'ed\'erale de Lausanne, CH-1015, 
  Switzerland}
\author{A.~Neronov}%
\affiliation{INTEGRAL Science Data Center, Chemin d'\'Ecogia 16, 1290 Versoix,
  Switzerland}
  
\author{O.~Ruchayskiy}%
\affiliation{Institut des
  Hautes \'Etudes Scientifiques, Bures-sur-Yvette, F-91440, France}
  
\author{M.~Shaposhnikov}%
\affiliation{\'Ecole Polytechnique F\'ed\'erale de Lausanne, CH-1015, 
  Switzerland}
\affiliation{CERN, Theory department, CH-1211 Geneve 23, Switzerland} %

\author{I.~Tkachev}%
\affiliation{CERN, Theory department, CH-1211 Geneve 23, Switzerland} %
\affiliation{\'Ecole Polytechnique F\'ed\'erale de Lausanne, CH-1015, 
  Switzerland}

\begin{abstract}
  We propose a strategy of how to look for dark matter (DM) particles
  possessing a radiative decay channel and derive constraints on their
  parameters from observations of X-rays from our own Galaxy and its dwarf
  satellites. When applied to the sterile neutrinos in keV mass range, it
  allows a significant improvement of restrictions to its parameters, as
  compared with previous works.
\end{abstract}

\maketitle

{\bf Introduction.} It was noticed long ago that a sterile neutrino with the
mass in the keV range appears to be a viable DM candidate~\cite{Dodelson:93}.
Moreover, being a \emph{warm} DM, sterile neutrino ease the tension between
observations and predictions of the \emph{cold} DM model on small scales.  The
interest to this scenario is revitalized since the discovery of neutrino
oscillations (see e.g.~\cite{Strumia:05} for review).  Indeed, one of the
simplest ways to explain this data is to add to the Standard Model several (at
least two) gauge singlet fermions -- right handed, or \emph{sterile}
neutrinos.  It has been demonstrated recently~\cite{Asaka:05a} that a simple
extension of the Standard Model by three singlet fermions with masses smaller
than the electroweak scale, dubbed the $\nu$MSM in~\cite{Asaka:05a}, allows to
accommodate the data on neutrino masses and mixings (with the exception of the
LSND anomaly), provides a candidate for DM particle in the form of sterile
neutrino, and allows to explain baryon asymmetry of the Universe. The
simplicity of the model, the similarity of its quark and lepton right-handed
sectors, together with a considerable number of phenomena it can
simultaneously describe, forces us to take this model seriously and thus
provides additional motivations for study of keV mass range sterile neutrino
as a DM candidate.

The sterile neutrino has a radiative decay channel, emitting a photon with
energy $E=m_s/2$ ($m_s$ being the mass of sterile neutrino).  Parametrically,
the decay width is proportional to $m_s^5\, \sin^22\theta $
\cite{Pal:81-Barger:95}, where $\theta$ is the mixing angle between active and
sterile neutrino.

If such a neutrino is a main ingredient of the DM, it is potentially
detectable in various X-ray observations. The most obvious candidates include
diffuse extragalactic X-ray background
(XRB)~\cite{Dolgov:00,Abazajian:01b,Mapelli:05,Boyarsky:05}; clusters of
galaxies~\cite{Abazajian:01b,Abazajian:05a,Boyarsky:06b};
galaxies~\cite{Abazajian:01b}, including our own.

The aim of the present Letter is to discuss the best strategy to search for a
DM sterile neutrino and to derive the constraints on its properties. Although
we concentrate on sterile neutrino, the constraints we get can be applied to
any DM candidate with a radiative two-body decay channel in a keV range. We
analyze various types of astrophysical objects and show that the strongest
constraints on sterile neutrino are coming from neutrino decays in the Milky
Way halo and in particular in the halo dwarf galaxies.  These objects were not
considered previously in this context.  The existing XMM Newton and HEAO-1
data allow us to improve over previous constraints, highlighting the potential
of new optimal cites for the searches of the signal from the sterile neutrino
decay.

{\bf Dark matter halo of the Milky Way (MW).} Energy flux produced by the DM
decay from a given direction into a solid angle $\Omega_\fov\ll 1$ is given by
\begin{equation}
\label{eq:8} 
  F =  \frac{\Gamma\Omega_\fov}{8\pi}\hskip -3ex\int\limits_{\text{line of 
      sight}} \hskip -3ex\rho_\dm(r)dr \;,
\end{equation}
where $\Gamma$ is the radiative decay width of sterile neutrino. To
determine the MW contribution into the flux~(\ref{eq:8}), one needs
to know the distribution of the DM in the halo.

Mass distribution within MW has been modeled by many authors.  Characteristics
which are relevant for our study are tightly constrained by the wealth of
detailed data available for this system.  As a reference we will choose the
mass distribution derived recently in~\cite{Klypin:02}, where physically
interesting models were selected by imposing additional constraints based on a
theory of halo formation. At large $r$ the halo density can be described by
the Navarro-Frenk-White (NFW) profile $\rho_{\textsc{nfw}}(r)
=\frac{M_\mathrm{vir}}{4\pi \alpha}\frac{1}{r(r_s+r)^2}$.  The MW halo
parameters of favorite models obtained in~\cite{Klypin:02} correspond to
$M_{\rm vir} = 1.0\times 10^{12} M_\odot$, $r_s = 21.5$ kpc and numerical
constant $\alpha\simeq 1.64$.

In the region of $r$ relevant for our study the halo density can be
also approximated by the isothermal profile
\begin{eqnarray}
\rho_{\rm halo} = \frac{v_h^2}{4\pi G_N}\; \frac{1}{r_c^2+r^2},
\label{density_profile_ITh}
\end{eqnarray}
where $v_h$ corresponds to contribution of the DM halo
into the Galactic rotation curve in its flat part, $v_h \approx 170$
km/s, see e.g.~\cite{Klypin:02}.

The NFW density profiles and (\ref{density_profile_ITh}) produce identical
fluxes from the direction of the Galactic anti-center if $r_c \approx 4$ kpc,
giving $ \int_{r_\odot}^\infty \rho_{\rm halo}\, dr \approx 0.7 \times
10^{22}\; {\rm GeV}/ {\rm cm}^{2}$. They also follow closely each other in the
range $3\; {\rm kpc} < r < 80\; {\rm kpc}$ (difference being less than 5\%).
Therefore, in estimates of the flux from directions which are outside of
$20^\circ$ circle around the Galactic center both NFW
and~(\ref{density_profile_ITh}) give the same results.

The halo density profile is less certain in the region $r < r_\odot$.  In
~\cite{Klypin:02} two distinct types of models were considered, with and
without exchange of angular momentum between DM and baryons. In the
model without momentum exchange the DM density profile at $r < 10$
kpc diverges even faster than NFW profile. In the model with angular momentum
exchange the DM density profile at $2\; {\rm kpc} < r < 10\; {\rm
  kpc}$ is less singular and rather resembles the isothermal sphere
Eq.~(\ref{density_profile_ITh}) with $r_c \approx 4$ kpc, but the DM
density is larger at $r < 2$ kpc as compared to the isothermal sphere.
Therefore, one can use Eq.~(\ref{density_profile_ITh}) with $r_c = 4 $~kpc as
a lower limit on the DM density when calculating fluxes from DM decays, and,
therefore, for putting a conservative bound on the sterile neutrino
mixing angle (e.g. the halo density at the Sun position in the model
Eq.~(\ref{density_profile_ITh}) is $0.25\;\mathrm{GeV/ cm}^{3}$, which is
smaller than the accepted value $0.3\;\mathrm{GeV/cm}^{3}$~\cite{pdg}).
Utilizing the NFW profile at all $r$ can only strengthen the bounds.

In the model Eq.~(\ref{density_profile_ITh}) the DM flux from the direction
$(b,l)$ (in galactic coordinates) into the solid angle $\Omega_\fov \ll 1$,
measured by an observer on Earth is given by
\begin{equation}
  \label{eq:19}
    F_\gal(\phi)\, {=}\, \frac {L_0}{R} \times \left \{ 
    \begin{array}{ll}
     \frac {\pi}{2} + 
        \arctan\left(\frac{r_\odot\cos\phi}
        {R}\right) , &   0\le\phi \le \frac{\pi}{2} \\
         \arctan\left(
  \frac{R}{r_\odot|\cos\phi|}\right), & \frac {\pi}{2}<\phi \le\pi
        \end{array} \right.,
\end{equation}
where $L_0\equiv\frac{\Gamma \Omega_\fov v_h^2}{32\pi^2 G_N}$, $R=\sqrt{r_c^2
  + r_\odot^2\sin^2\phi}$, and $\cos\phi =\cos b\,\cos l$.  
For example, ${F_\gal(90^\circ)}/{F_\gal(180^\circ)} \simeq 1.52$.

{\bf Search for a preferred observation.} Let us compare the Galaxy
contribution to the DM decay flux, computed above,
with those of other astrophysical objects.

\emph{(i) XRB.} Although the DM has a very narrow radiative decay width, the
cosmological DM decay contribution to the XRB gets significantly broaden due
to the contributions from various red shifts.  The resulting differential flux
for $E < m_s/2$ is given by
\begin{equation}
  \label{eq:5}
  \frac {d^2 F_\xrb}{dE\,d\Omega} = \frac{\Gamma}{4\pi
    H_0 m_s}\frac{\rho_\dm^0 (2E)^{3/2}}{\sqrt{8 E^3 \Omega_\Lambda + \Omega_m
      m_s^3}} ~,
\end{equation}
where $\rho^0_\dm = 1.2\times 10^{-6}\,\mathrm{GeV/cm^3}$ is the average DM
density in the Universe, $\Omega_\Lambda$ and $\Omega_m$ are the cosmological
constant and matter contributions to the density of the Universe.  The paper
\cite{Dolgov:00} looked at restrictions from XRB, assuming that DM is uniform
up to very small $z$.  This question was further addressed
in~\cite{Abazajian:01b,Mapelli:05} and finally the most stringent constraint
in the $(\sin^22\theta,m_s)$ plane from XRB was obtained recently in
\cite{Boyarsky:05}.

One can compare flux~(\ref{eq:5}), integrated over all $E$ with the
galactic contribution: $ F_\gal/F_\xrb = \mathcal{R}\, r_c H_0 $, where
$\mathcal{R}\sim \rho^0_\gal/\rho^0_\dm\sim 10^6$ is \emph{overdensity} of the
Galaxy as compared to the average density of DM in the universe. With $r_c
\sim 4\kpc$, one arrives to the conclusion that $F_\gal/F_\xrb \sim 1$, i.e.
Galactic contribution is \emph{comparable} with the total DM decay flux from
all red shifts.

However, for a modern X-ray instrument with good spectral resolution $\Delta
E\ll m_s$ (e.g. XMM-Newton) one should compare contributions from the Galaxy
and from uniform cosmological distribution into XRB within $\Delta E$.  The
ratio $F_\gal/F_\xrb$ then gets enhanced by the factor $E/\Delta E$ which is
$10\div50$ for EPIC cameras on board of XMM-Newton, i.e. for XMM the Galactic
signal is 1 to 2 orders of magnitude \emph{stronger} than the
contribution from uniform distribution of DM in the universe.

\emph{(ii) Clusters.} Let us now analyze the flux enhancement from Coma and
Virgo clusters of galaxies~\cite{Abazajian:01b,Boyarsky:06b}. The DM
distribution in these clusters can be approximated by \emph{isothermal
  $\beta$-model}~\cite{Cavaliere:76,Sarazin:77}
\begin{equation}
  \label{eq:25}
  \rho_\cl(r) =
  \rho^0_\cl\frac{3+(r/r_\cl)^2}{(1+(r/r_\cl)^2)^2} \;.
\end{equation}
Integrating it according to Eq.~(\ref{eq:8}) and using the fact that
the Coma cluster is located perpendicularly to the Galactic plane,
and has core radius $r_\cl \simeq 0.3\mpc$ and central density $
\rho^0_\coma = 
10^{-2}\,\mathrm{GeV/cm^3}$ (see e.g.~\cite{Boyarsky:06b} for
discussion), we get
$F_\gal(90^\circ)/{F_\coma}\approx 0.25$.  Similar estimate
for the center of the Virgo cluster ($\rho^0_\virgo \simeq
2.3\,\mathrm{GeV/cm^3}$, $r_\cl \approx 10$~kpc) shows that Galactic
contribution is $\sim 10\% $. Therefore, one could conclude that clusters are
preferable objects for DM detection \cite{Abazajian:01b,Abazajian:05a}.  

These results have been recently reanalyzed by~\cite{Boyarsky:06b},
where it was shown that the actual restriction from clusters is only
by about a factor of 3--4 (in $\theta^2$ for given $m_s$) better than
those from XRB.  The reason for such a modest improvement is the
following.  The detection of the DM decay line in galaxy clusters is
complicated by the fact that most of them show strong emission
precisely in keV range.  Indeed, the virial theorem immediately tells
us that the temperature of the intercluster medium is $
T_\mathrm{gas}\sim G_N\,m_p\,{\cal R}\rho^0_{\dm}d^2$, where $d$ is
characteristic size.  For overdensity $\mathcal{R}\sim 10^3$ and size
$d\sim1$~Mpc, the temperature $T_\mathrm{gas}$ is always in the keV
range, which makes it hard to detect a DM decay line
against a strong X-ray continuum.

\emph{(iii) Dwarf galaxies.} There should be an enhancement of the flux in
directions of dwarf spheroidal galaxies, which are satellites of the MW.
Promising satellites with large mass to light ratio are Draco and Ursa Minor.
Density profiles of both galaxies can be modeled by the isothermal sphere with
${v_h \approx 22}$ km/s and $r_c \approx 100$ pc \cite{Wilkinson:06}. This
gives for the contribution to the flux from the dwarf galaxy (along the line
which passes through the core of the satellite) $\int \rho\, dr \approx 3.3
\times 10^{22}\; {\rm GeV}/{\rm cm}^{2}$.

Contribution of the Galaxy halo flux in directions of both satellites $\approx
1.0 \times 10^{22}\; {\rm GeV}/ {\rm cm}^{2}$.  Therefore, in directions of
both satellites the 4-fold local enhancement of the flux is expected, while
the total flux matches the flux from the central region of the MW and the flux
from clusters.  The advantage of observing the dwarf satellites, as compared
to clusters or to the Galactic center, is lower level of X-ray background
contamination and clear signature of the signal, namely, local flux
enhancement within single field of view of X-ray telescope. 

\emph{Therefore}, we see that the local DM halo (especially dwarf satellite
galaxies) can provide \emph{the strongest restriction} on the parameters of
the sterile neutrino as DM candidates.

\begin{figure}[t]
  \centering %
  \includegraphics[width=\linewidth]{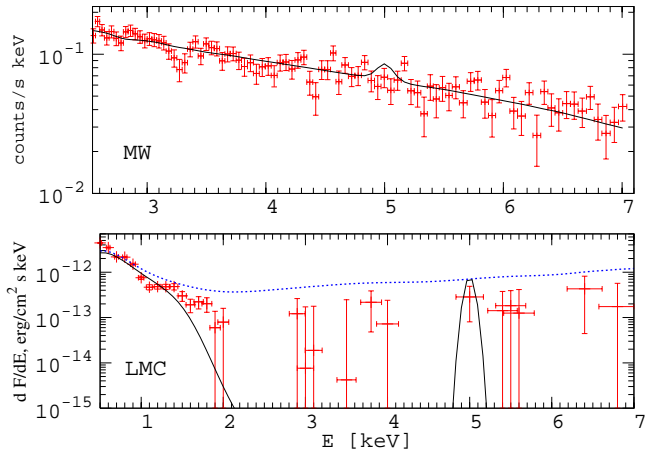} %
  \caption{Upper panel: Method used to obtain restrictions on DM decay in MW
    from blank sky XRB observations.  The data is fitted by a power law
    (reduced $\chi^2 = 1.07$ for 82 d.o.f.)  and \texttt{XSPEC v11.3.2} is
    used to put a $3\sigma$ limit on the presence of DM line (via command
    \texttt{"error <line norm> 9.0"}).\\ Lower Panel: Method used to obtain
    restrictions from LMC. Flux rapidly decreases for $E\gtrsim 2$~keV, most
    of the data points at higher energies are zero within statistical
    uncertainty.  The solid green line is the sum of the total flux plus
    $3\sigma$ per energy bin. At $E\gtrsim 2$~keV the bound is dominated by
    errors and therefore can be improved significantly by increasing
    statistics.  } %
  \label{fig:method}
\end{figure}

\begin{figure}[t]
  \centering %
  \includegraphics[width=\linewidth]{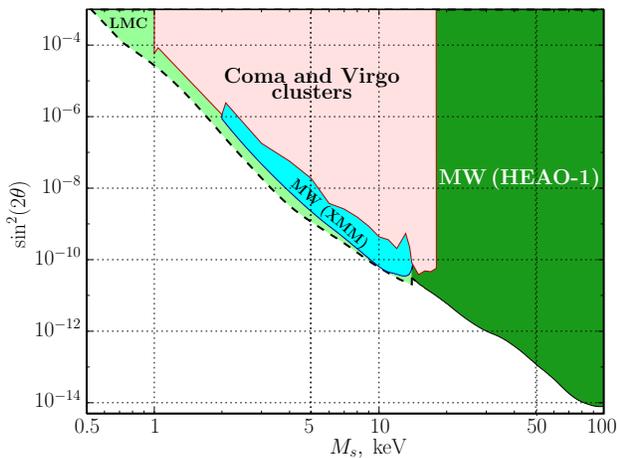} %
  \caption{3$\sigma$ restrictions on DM decay line from Milky way  halo (XMM
    and HEAO-1 observations).  Dashed line: total flux + $3\sigma$ restriction
    from LMC.  Also shown the previous strongest limit from the clusters of
    galaxies~\cite{Boyarsky:06b}.
} %
  \label{fig:exclusionFromLMC}
\end{figure}

{\bf Restrictions from local halo DM contribution.}  

We have analyzed XMM blank sky observations of~\cite{Read:03} (exposure time
$\sim 200$~ksec), taking into account MW contribution and putting a $3\sigma$
bound on the DM line flux in the region $1\kev < E < 7\kev$
(FIG.~\ref{fig:method}). The
exclusion plot in the region $m_s > 6\kev$ was obtained
by using the HEAO-1 measurements of XRB~\cite{Gruber:99}. As spectral
resolution of HEAO-1 is about 25\%, the resulting correction due to MW is not
so drastic as compared to the results of Ref.~\cite{Boyarsky:05}.  

Unfortunately, X-ray observations of Draco and Ursa Minor dwarfs are not
available currently.  However, approximately the same signal is expected from
the Large Magellanic Clouds (LMC) albeit with larger uncertainties.  Using
isothermal sphere, with the following halo parameters, ${v_h \approx 50}$ km/s
and $r_c \approx 1$ kpc \cite{Marel}, we obtain $\int \rho\, dr \approx 2.8
\times 10^{22}\; {\rm GeV}/{\rm cm}^{2}$ for the LMC contribution, while MW
halo contribution in the direction of LMC is again $\approx 1.0 \times
10^{22}\; {\rm GeV}/{\rm cm}^{2}$ (the use of the NFW profile gives even
larger value for flux).

Therefore, as an illustration we have processed one of the observations of the
LMC (XMM obs ID: 0127720201, exposure $\sim 20$~ksec).  Subtracting the blank
sky background~\cite{Read:03}, one sees that the flux is zero within
statistical uncertainty for $E\gtrsim 2$~keV.  Similar reduction of the
background is expected for dwarf satellites. In this situation we put an upper
limit on the flux of the DM decay line, by demanding it to be smaller than
total flux plus its $3\sigma$ error in an energy bin equal to spectral
resolution, see Fig.\ref{fig:method}.  One can see that increasing the
exposure time, one can significantly lower the restriction especially in the
region $E \gtrsim 2$~keV.

{\bf Discussion.} In this paper, we have shown that the best objects for the
search of the DM with radiative decay channel is the MW halo, including dwarf
satellite galaxies (e.g.  Draco and Ursa Minor).  To illustrate this, we put
the strongest restrictions on parameters of sterile neutrino \emph{(i)}
searching for the MW DM decay signal in the blank sky XRB; \emph{(ii)} using
LMC observations.  One can see that improving statistics one can significantly
improve bounds from such objects. Our analysis also shows that improvement of
the spectral resolution of X-ray instruments (even by means of decreasing
imaging capabilities) is crucial to continue the search of DM decay line.

Of course, all constraints on the sterile neutrino mixing angle, derived from
X-ray observations, suffer from the uncertainties in the DM profiles.
However, in our analysis we tried to be as conservative as possible. Namely,
we present only the results coming from the study of directions which are away
of the Galactic center. In this case the most relevant parameter, which
influences the line of sight integral Eq.  (\ref{eq:8}) is $v_h$,
corresponding to the contribution of the DM halo to the Galactic rotation
curve at large distances, away form the core. The total (DM plus baryons)
rotational curve is measured directly. Although the subtraction of the baryon
contribution is model dependent, this dependence is weakest for the directions
opposite to the Galactic center. 

Moreover, we have found, quite remarkably, that the line of sight integral
Eq.~(\ref{eq:8}) is roughly the same for all studied DM dominated objects,
from {cosmological background} to clusters of galaxies to dwarfs
satellites.  
This suggests that the
expected DM decay signal should be roughly the same for all of them.  Better
constrains are obtained for the objects whose X-ray background is lower.  This
makes dwarf satellites more suitable as compared to clusters. In addition, in
the most DM dominated satellites, such as Draco and Ursa Minor, not only the
X-ray background is lower, but uncertainties due to subtraction of baryonic
component from the galactic rotational curve are also minimized. The
restrictions, based on the different (types of) objects, further minimize
these uncertainties.

The limits we derived here allow to strengthen the bounds on sterile neutrinos
in different models of particle physics, shed light on the possible mechanisms
of their production in the early universe, and constrain different
astrophysical phenomena that might be related to their existence.

(i) In the Standard Model with addition of just one sterile neutrino (assuming
the absence of sterile neutrinos above the temperature $\sim 1$ GeV and charge
neutrality of the plasma) the relic abundance of sterile neutrinos can be
expressed through $m_s$ and $\theta$ \cite{Dodelson:93}. This relation (quite
uncertain, since the sterile neutrinos are mainly produced at temperatures
${\cal O}(150)$ MeV, exactly where the description of the strongly interacting
plasma is most complicated \cite{Asaka:06b}) allows one to find, potentially,
an upper limit on the sterile neutrino mass in this particular scenario.
Taking as a rough estimate the computation of \cite{Abazajian:05a} and our
X-ray constraints we arrive to an upper bound $m_s \lesssim 3$ keV. This may
be contrasted with the lower bound on the mass of sterile neutrino (derived in
the same model with the same assumptions) coming from the analysis of the
Ly-$\alpha$ forest data \cite{Hansen:01}: $m_s > 2$ keV~\cite{Viel:05}, $m_s >
1.7 $ keV \cite{Abazajian:05b} leaving a very limited allowed mass range for a
sterile neutrino. If a more recent result of~\cite{Seljak:06}, $m_s > 14$ keV
is proven to be correct, and uncertainties related to poor knowledge of QCD at
the relevant range of temperatures happen to be not substantial, this scenario
will be ruled out by cosmological and astrophysical observations. The same
conclusion is true \cite{Asaka:06} for the $\nu$MSM if the same assumptions
about the initial state are taken for granted. This would make the production
mechanisms of the sterile neutrinos related to
inflation~\cite{Shaposhnikov:06} or large lepton asymmetries \cite{Shi:98}
more important.

(ii) The X-ray observations allow to predict the masses of active
neutrinos in the $\nu$MSM. In  \cite{Boyarsky:06a} was shown, that
the XRB limits of \cite{Boyarsky:05} imply that the  lightest active
neutrino must have a mass $m_\nu < 3\times 10^{-3}$, provided
$m_s>1.8$ keV. In case of normal hierarchy two other masses are given
by the observed mass square differences $\sqrt{\Delta m^2_{\rm
solar}}\simeq (8.5-9.5)\times 10^{-3}$ eV and $\sqrt{\Delta m^2_{\rm
atm}}\simeq 0.04-0.06$ eV. (In the case of inverted hierarchy masses
of both neutrinos are $\sqrt{\Delta m^2_{\rm atm}}$). With the
improved bound, derived in this paper, this result is true for even
smaller sterile neutrino masses, $m_s > 1.3$ keV.

(iii) Our constraints, combined with those of \cite{Seljak:06}, put
severe bounds on explanation of pulsar kick velocities by dark matter
sterile neutrinos, proposed in 
\cite{Kusenko:97-Fuller:03-Barkovich:04} and on the mechanism of
early reionization by radiative decays of sterile neutrinos,
suggested in  \cite{Biermann:06}.

{\bf Acknowledgements.} We thank A.Kusenko and M.Markevitch for discussions.
MS was supported in part by the Swiss Science Foundation. OR was supported by
a \emph{Marie Curie International Fellowship} within the FP-6 Programme.

{\bf Note added.} After finishing this paper we became aware of an independent
analysis, where the possibility of measuring decaying DM particles
from our Galaxy, using Chandra blank sky data, was considered
\cite{Hansen:06}. The conclusions, reached in that paper on MW, are similar to
ours.


\end{document}